\documentstyle[12pt,aps,preprint]{revtex}

\begin{document}

\newcommand{\be}{\begin{equation}}
\newcommand{\ee}{\end{equation}}
\newcommand{\bea}{\begin{eqnarray}}
\newcommand{\eea}{\end{eqnarray}}

\title{Making Baryons Below the Electroweak Scale}

\author{Mark Trodden}

\address{Physics Department, Case Western Reserve University, 10900 Euclid 
Avenue, Cleveland, OH 44106-7079, USA\\E-mail: trodden@huxley.cwru.edu}

\maketitle

\begin{abstract}
I describe a new way for baryogenesis to proceed, which evades many of
the problems of GUT and electroweak scenarios.
If the reheat temperature after inflation is below the electroweak scale,
neither GUT baryon production nor traditional electroweak baryogenesis can
occur. However, non-thermal production of sphaleron configurations via 
preheating could generate the observed baryon asymmetry of the universe.
Such low scale baryon production is particularly attractive since it evades
a number of strong constraints on reheating from gravitino and moduli
production.
\end{abstract}

\setcounter{page}{0}
\thispagestyle{empty}

\vfill

\noindent CWRU-P2-00\hfill 

\vfill
\eject

\baselineskip 20pt plus 2pt minus 2pt

\section{Introduction}
Over the past twenty to thirty years, a variety of microphysical explanations
for the observed baryon asymmetry of the universe have been proposed.  
Initially. Grand Unified Theories (GUTs), which easily satisfy
Sakharov's three criteria for baryogenesis,
were demonstrated to be able to account for the observed baryon to
photon ratio in the Universe today \cite{GUTs}.  Although proton
decay experiments soon ruled out the simplest theories, GUT
baryogenesis remained a viable possibility in more complicated
models. However, GUTs also lead to several cosmological problems. Since 
inflation erases any preexisting asymmetry, GUT baryogenesis is only possible
if the reheating scale following inflation is large.
It has been realized for some time that this raises the possibility of 
unacceptable defect and, in supersymmetric (SUSY) models, gravitino and moduli,
production after inflation \cite{gravitinos}. However, as we've heard in a 
number of talks here \cite{gravitinotalks}, these concerns have 
become much more pressing recently, with the realization that, in the context 
of preheating, gravitino and moduli production can be so efficient as to 
constrain the reheat temperature to be less than $10$ GeV in some models. 

An additional problem for GUT baryogenesis contained
the seeds for potentially viable
baryogenesis at the much lower electroweak scale ($\sim 10^2$ GeV).  
Coherent configurations
of electroweak gauge and Higgs fields, 
first pointed out by 't Hooft \cite{tHooft76}, can violate baryon
number via non-perturbative physics.   At zero temperature this effect is
exponentially suppressed by the energy of a field configuration called the
{\it sphaleron}, and is essentially irrelevant.  However, as pointed
out by Kuzmin, Rubakov and Shaposhnikov \cite{krs}, and later discussed by
Arnold and McLerran \cite{a&m}, at finite temperature, sphaleron
production and decay can be rampant.  This has the virtue of allowing
copious baryon number violation, but can also be a
curse.  If the universe remains in thermal equilibrium until sphaleron
production ceases, the net effect of these processes will be to drive the
baryon number of the universe to zero, unless careful precautions are
made to ensure either out of equilibrium sphaleron decay, or quantum
number restrictions which forbid the elimination of the net baryon number.
In the light of recent lattice and experimental data, this seems to require
new fields at the weak scale, perhaps those predicted by SUSY.

Finally, once again, if the gravitino reheating constraint is sufficiently
strong, we can not allow the universe to reheat to a sufficient temperature
to allow even electroweak baryogenesis to take place.

Thus, thermal sphaleron production creates both challenges and opportunities 
for the generation of the baryon asymmetry.  While it can wipe out any 
baryon number generated at the GUT scale, it offers the possibility
of electroweak baryogenesis, although in practice this is quite difficult to
achieve (for reviews see \cite{reviews}).

As I alluded to earlier, {\it reheating} after cosmological inflation has been
carefully rethought over the last few years. Studies of the inflaton 
dynamics have revealed the possibility of a
period of parametric resonance, prior to the usual scenario of energy
transfer from the inflaton to other  fields. This phenomenon, which is
characterized by large amplitude, non-thermal excitations in both the
inflaton and coupled fields, has become known as {\it preheating}
\cite{KLS 94,STB 95}. Two particularly interesting consequences of this
are the strict graviton and moduli constraints that result 
\cite{gravitinotalks}, and the idea that
topological defects may be produced after inflation
even when the final reheat temperature is lower than the symmetry breaking
scale of the defects \cite{KLS 94,KKLT,kofm,tch}. 

In this article, I describe how all these ideas can be combined
to yield a viable and attractive model which obviates many of the
problems with both standard GUT, and electroweak baryogenesis 
\cite{K&T 99}.  In particular, I show how baryogenesis might still occur, 
even if inflation ends with reheating below the electroweak scale.

\section{The Basic Mechanism}
The fundamental idea is that, if topological defects can be
produced non-thermally during preheating, then so can coherent 
configurations of gauge and Higgs fields, carrying nontrivial 
values of the Higgs winding number
\be
N_H(t) = \frac{1}{24\pi^2} \int d^3x\, \epsilon^{ijk} \hbox{Tr}
[U^{\dagger}\partial_iUU^{\dagger}\partial_jUU^{\dagger}\partial_kU] \ .
\label{higgswinding}
\ee
In this parameterization, the $SU(2)$ Higgs field $\Phi$ has been expressed as
$\Phi = ({\sigma}/{\sqrt{2}}) U  $, where $\sigma^2 =
2\left(\varphi_1^*\varphi_1 + \varphi_2^*\varphi_2 \right) 
= {\rm Tr} \Phi^\dagger
\Phi$,  and $U$ is an $SU(2)$-valued matrix that is uniquely defined
anywhere $\sigma$ is nonzero. 

These winding configurations are not stable and evolve to a vacuum 
configuration plus radiation. In the process fermions may be 
anomalously produced. If the fields relax to the vacuum by changing the Higgs 
winding then there is no anomalous fermion number production. However, if 
there is no net change in 
Higgs winding during the evolution (for example $\sigma$ never
vanishes) then there is anomalous fermion number production.
Since winding configurations will be produced
out of equilibrium (by the nature of preheating) and since CP-violation 
affects how they unwind, all the ingredients to produce a baryonic 
asymmetry are present (see \cite{LRT 97} for a detailed discussion of
the dynamics of winding configurations).

If the final reheat
temperature is lower than the electroweak scale, then then production of
small-scale winding configurations by resonant effects is analogous
to the production of local topological defects. In fact, the configurations 
that are of interest can be thought of as {\it gauged textures}.

Given this connection, a rough underestimate of the number 
density of winding configurations may be obtained by counting 
defects in recent numerical simulations
of defect formation during preheating \cite{KKLT}, while keeping 
in mind that the important case
is when the symmetry breaking order parameter is not the
inflaton itself, but is the electroweak $SU(2)$
Higgs field, and is coupled to the inflaton. The relevant quantity is
the number density of defects directly after preheating, since 
winding-anti-winding pairs of configurations will not typically
have time to find each other and annihilate before they decay.  
Finally, since the Higgs
winding is the only non-trivial winding present at the electroweak scale,
it is reasonable to assume that any estimates of defect production in 
general models can be quantitatively carried over to estimate of the
relevant Higgs windings for preheating at the electroweak scale.

\section{A (Too?) Simple Example}
Before I make an estimate of the baryon asymmetry from this mechanism,
I'll provide an example of a toy model which satisfies all the relevant
constraints.

Consider the potential
\be
V(\phi,\chi) = \frac{1}{2} m^2 \phi^2 + \frac{1}{2}g^2\phi^2 \chi^2
+ \frac{1}{4} \lambda (\chi^2 - \chi_0^2)^2 \ ,
\ee
for an inflaton $\phi$, coupled to the electroweak Higgs field $\chi$. 
\footnote{This
model has also been independently proposed in a similar context in
\cite{guys}, and for a description of this see Misha Shaposhnikov's 
contribution to these proceedings \cite{misha}}
Here $\chi_0 = 246$ GeV is the electroweak symmetry breaking scale, 
$m$ is the (false vacuum) inflaton mass, 
and $\lambda$ (the Higgs self-coupling, here assumed to be of order unity) 
and $g$ are dimensionless constants.

The mechanism only works if parametric resonance into electroweak
fields occurs in this model. The condition for this to happen is
\cite{klebtkach}
\be
q = \frac{g^2 \phi_0^2}{2m_{\phi}^2} > 10^3 \ ,
\label{qcond}
\ee 
where $\phi_0$ is the value of $\phi$ at the end of inflation.
For the values quoted here, this condition 
yields $g<10^{-2}$ (I'll take $g\sim 10^{-2}$).
It is important that the temperature fluctuations in the cosmic microwave 
background (CMB), given by 
\be
\frac{\delta T}{T} \sim g \frac{\chi_0^5}{M_p^3 m^2} \ ,
\ee
for the values I've chosen here,
are of the correct magnitude. Clearly this is satisfied by the choice
$m \sim 10^{-21}$ GeV.
Finally, since the reheat temperature in this model is roughly bounded by
$T_{RH} \leq (m\phi_0)^{1/2}$, the requirement that any baryons produced 
not be erased by equilibrium sphaleron processes is also satisfied.

This is not a particularly natural toy model, and in fact, it may develop 
problems if we go beyond tree level \cite{david}. However, the point of this 
example is merely to provide an existence proof which makes explicit the 
constraints on such a possibility. 

\section{Calculating the Asymmetry}
Consideration of topological defect production following inflation has
been discussed by several authors {\cite{kofm,tch}.
For definiteness, let us focus on the results of 
Khlebnikov {\it et al.}. These show that, for
sufficiently  low symmetry breaking scales, the {\it initial} number
density of defects produced is very high.  Here, by initial, I mean 
the number seen after copious  symmetry-restoring transitions cease. One may
perform an estimate from the first frame of Figure~6. of reference
\cite{KKLT}. The box size has physical size $L_{phys} \sim 50 \eta^{-1}$
where $\eta$ is the symmetry breaking scale, and I've
assumed couplings of order unity. In this box there are of order $N=50$
defects at early times. Thus, a rough estimate of
the number density of winding configurations is
\be
n_{\rm configs} \sim \frac{N}{L_{phys}^3} \sim 4 \times 10^{-4} \eta^3\ .
\ee

In order to make a simple estimate of the baryon number
produced, it remains to show how CP-violation may bias the decays of 
these configurations to create a net baryon excess.

The effect of CP-violation on winding configurations can be very 
complicated, and in general depends strongly on the shapes of the 
configurations \cite{LRT 97} and the particular type of CP-violation. 
However, in general, the situation considered here, when out of 
equilibrium configurations are produced in a background low-temperature 
electroweak plasma most closely
resembles local electroweak baryogenesis in the ``thin-wall'' regime. Winding
configurations are produced when non-thermal 
oscillations take place in a region of space and restore the symmetry
there. Since
the reheat temperature is lower than the electroweak scale, as the region 
reverts rapidly to the low temperature phase, the winding configuration is
left behind. 
In the absence of CP-violation in the coupling of the inflaton to the
standard model fields, 
a CP-symmetric ensemble 
of configurations with $N_H=+1$ and $N_H=-1$ will be produced.
(i.e. the probability for finding
a particular $N_H=+1$ configuration in the ensemble
is equal to that for finding its CP-conjugate $N_H=-1$ configuration.)
Then, without electroweak CP- violation, for every $N_H=+1$
configuration which relaxes in a baryon producing fashion
there is an $N_H=-1$ configuration which produces anti-baryons, and no
net baryogenesis occurs.
However, with CP-violation there will 
be some configurations which produce baryons whose CP-conjugate 
configurations relax without
violating baryon number.

While an analytic computation of the effect of CP-violation does not exist 
\cite{LRT 97}, there exist numerical simulations (e.g. \cite{Moore}), from 
which one expects that the asymmetry in the number density of
decaying winding configurations should be proportional to a dimensionless
number, $\epsilon$, parameterizing the strength of the source of CP-violation.
Now, at the electroweak scale the entropy density is 
$s\simeq 2\pi^2 g_* T^3/45$, where $g_*\sim 100$ is the effective number
of massless degrees of freedom at that scale. Thus, the final baryon to
entropy ratio generated is
\be
\eta \equiv \frac{n_B}{s} \sim \epsilon \, g_*^{-1} 
\frac{n_{\rm configs}}{T_{RH}^3} \ .
\label{etafinal}
\ee
Plugging in the approximate numbers obtained earlier, this yields
\be
\eta \equiv \sim 10^{-6} \epsilon \ .
\ee
This is the final estimate.

This estimate is quite rough, and the explicit model
presented is merely a toy model.  However these suggest that the
mechanism proposed here could viably result in a
phenomenologically allowed value of $\eta
\sim 10^{-10}$, with CP violating physics within the
range predicted in SUSY models for example.

\section{Conclusions}
I have described a new mechanism for baryogenesis, that is effective below
the electroweak scale.
The primary advantages of such a mechanism are that 
no thermal sphaleron production subsequently takes place
to wash out any baryon number that is produced, and that no excess production
of gravitinos or monopoles occurs, evading a very strong (although 
model-dependent) constraint. A more complete analysis of the mechanism 
requires a numerical solution to the coupled
$SU(2)$-inflaton equations of motion, in the presence of CP-violation.

\section*{Acknowledgements}
I would like to thank Lawrence Krauss for a stimulating and enjoyable 
collaboration on this project. I also thank Matthew Parry, Richard Easther and 
Lisa Randall for helpful discussions. Finally, many thanks to the organizers, 
particularly Goran Senjanovic and Rachel Jeannerot for all their hard work, 
for making the conference fun, and for the excellent coffee.
This work was supported by the US Department of Energy.

\end{document}